\documentclass[12pt]{article}

\usepackage[ansinew]{inputenc}
\usepackage[english]{babel}
\usepackage{amsmath,amsfonts,amssymb,mathrsfs,cancel}
\usepackage[dvips]{graphicx}

\usepackage{epsfig}
\usepackage{graphicx}
\usepackage{subfigure}
\usepackage{cite}
\begin{document}

\def\Xint#1{\mathchoice
{\XXint\displaystyle\textstyle{#1}} 
{\XXint\textstyle\scriptstyle{#1}} 
{\XXint\scriptstyle\scriptscriptstyle{#1}} 
{\XXint\scriptscriptstyle\scriptscriptstyle{#1}} 
\!\int}
\def\XXint#1#2#3{{\setbox0=\hbox{$#1{#2#3}{\int}$ }
\vcenter{\hbox{$#2#3$ }}\kern-.5\wd0}}
\def\ddashint{\Xint=}
\def\dashint{\Xint-}

\newcommand{\be}{\begin{equation}}\newcommand{\ee}{\end{equation}}
\newcommand{\bea}{\begin{eqnarray}} \newcommand{\eea}{\end{eqnarray}}
\newcommand{\ba}[1]{\begin{array}{#1}} \newcommand{\ea}{\end{array}}

\long\def\symbolfootnote[#1]#2{\begingroup%
\def\thefootnote{\fnsymbol{footnote}}\footnote[#1]{#2}\endgroup} 

\numberwithin{equation}{section}

\def\a{\alpha }
\def\p{\partial }
\def\g{\gamma}
\def\w{\omega }
\def\lb{\lambda}
\def\lbb{\lambda^ 2}
\def\non{\nonumber}
 \def\vareps{\varepsilon }


\rightline{March, 2012}

\bigskip

\begin{center}

{\Large\bf  
A note on perturbation series in \\
  supersymmetric gauge theories
}

\bigskip
\bigskip

{\it \large  Jorge G. Russo\symbolfootnote[1]{On leave of absence from Universitat de Barcelona and Institute of Cosmos Sciences, Barcelona, Spain.}}
\bigskip

\end{center}

{\it

\begin{center}

  Perimeter Institute for Theoretical Physics,
Waterloo, Ontario, N2L 2Y5, Canada

\medskip

 Instituci\'o Catalana de Recerca i Estudis Avan\c cats (ICREA)\\
Pg. Lluis Companys, 23, 08010 Barcelona, Spain

\end{center}

}
\bigskip
\bigskip

\begin{abstract}

Exact results in supersymmetric Chern-Simons and ${\cal N}=2$  Yang-Mills theories
 can be used to examine the quantum behavior of observables and the structure of the perturbative series.
For the $U(2)\times U(2)$ ABJM model, we determine the asymptotic behavior of the perturbative series for the partition function 
and write it as a Borel transform. 
Similar results are obtained for  ${\cal N}=2$ $SU(2)$ super Yang-Mills theory with four fundamental flavors
and in ${\cal N} =2^*$ super Yang-Mills theory, 
for the partition function as well as for the expectation values for Wilson loop and 't Hooft loop operators 
(in the 0 and 1 instanton sectors).
In all examples, one has an alternate perturbation series where the coefficient of the $n$th term  increases as $n!$, and 
 the perturbation series are Borel summable.
We  also calculate the expectation value for a  Wilson loop operator in the ${\cal N} =2^*$ $SU(N)$ theory at large $N$
in different regimes of the 't~Hooft gauge coupling and  mass parameter. For large masses, the calculation reproduces
the running gauge coupling for the pure ${\cal N} =2$ SYM theory.

\end{abstract}

\clearpage

\tableofcontents

\section{Introduction}

An intriguing feature of quantum field theory is that the number of Feynman diagrams of $n$th order
increases as $n!$ \cite{Dyson:1952tj,Lipatov:1976ny}. From early times, this led to the belief that  perturbation series is divergent.
In some cases it may be rendered convergent by the extensively studied technique of  Borel transformation \cite{thoofterice}.
If one has the series
$$
Z(g)=\sum_{n=0}^\infty c_n\ g^n\ ,
$$
by inserting a ``1",
$$
1= {1\over n!} \int_0^\infty dt \ e^{-t}\ t^n\ ,
$$
and commuting integral and sum, one finds a new expression
\be
Z(g)= \int_0^\infty dt\ e^{-t} B(t g)\ ,\ \ \ \ B(x)=\sum_{n=0}^\infty {c_n\over n!}\  x^n\ ,
\label{boro}
\ee
where the expansion series for $B(x)$ has better convergence properties.
In particular, if $c_n\sim n!$ then $B(x)$ will have finite radius of convergence.
In this case $B(x)$ will have at least one singularity somewhere in the complex (Borel) plane at some finite $x$.
As long as this singularity does not lie on the real positive axes, the integral over $t$ can be performed,
obtaining a finite expression for $Z(g)$ out of the original divergent expression.
The trick of course lies in the illegitimate step of commuting the infinite sum  and the integration;
although it may appear as an artificial trick, the examples considered in this paper give
 a new  justification in support of the Borel resummation procedure.

Some singularities of the Borel transform $B(x)$ are associated with instanton solutions 
\cite{Lipatov:1976ny}. The special thing about instantons is that they tend to produce singularities on the real axes of $B(x)$.
In non-abelian gauge theories, instantons may produce singularities on the negative real axes, in such a case they are harmless.
Configurations of instantons and anti-instantons of vanishing topological charge  in current-current correlation functions
are believed to produce singularities on the positive real axes.
There can also be singularities which lie away from the real axes, and these typically correspond to (unphysical) complex field configurations.
In non-abelian gauge theories the main 
problem in applying the Borel transform arises from renormalons \cite{thoofterice}, associated with terms in the operator product expansion. They can produce singularities on the positive real axes, leading to an ill-defined Borel integral \cite{Lautrup:1977hs}.
This obstruction has been extensively studied (see e.g. \cite{Marino:2008ya} and references therein) leading to the theory of resurgent functions and there are prescriptions to construct well-defined
quantities by deforming the contour. This gives rise to an imaginary part  which in some cases  describes a physical instability of the system (an example is the
 ground state energy of the cubic anharmonic oscillator).

In non-abelian gauge theories in four dimensions, many of these considerations are incomplete insofar  as one does not know in general closed analytic expressions to all orders in the coupling
for any process.  In some cases, even obtaining partial results  require a complicated analysis.
However, the recent exact results in supersymmetric gauge theories obtained by localization techniques open the possibility of
revisiting these old issues and understanding in a clean setting the origin of perturbation theory divergences.

${\cal N}=2$ supersymmetric gauge theories have a rich quantum dynamics which in many respects reproduces the expected physics of QCD.
The exact low energy effective action constructed by Seiberg and Witten \cite{Seiberg:1994rs,Seiberg:1994aj} shows chiral symmetry breaking and quark confinement in the theory deformed by ${\cal N}=1$ preserving mass terms, via condensation of magnetic monopoles. More recently, exact results have been obtained 
for the partition function and for the expectation value of gauge invariant operators using localization \cite{Pestun:2007rz,Gomis:2011pf}.
 Implementing localization requires the existence of a fermionic symmetry. One can show that the functional integral reduces to a submanifold of field configurations invariant under the fermionic symmetry
that solve the saddle point equations. The complete effect of quantum fluctuations is then captured by the one-loop determinant.

In ${\cal N}=4$ SYM the one-loop determinant is absent and as we will see this has a striking effect on the behavior of the perturbative series.
One exact result  in ${\cal N}=4$ $SU(N)$ SYM is the expectation value of the circular Wilson loop operator obtained by Drukker and Gross \cite{Drukker:2000rr},
extending the result of \cite{Erickson:2000af} beyond the planar limit.
The result is expressed in terms of the following closed analytic expression:
\be
\langle W(C_{\rm circle}) \rangle =\frac{2e^{-{g^2(1+N)\over 8N}}}{N! g}
\int_0^\infty dt\ e^{-t} t^{N-{1\over 2}} I_1(\sqrt{t} g )\ ,
\label{druk}
\ee
where $I_1$ is the modified Bessel function.
Comparing with (\ref{boro}), we see that the integral is already given as a ``Borel transform",
\be
\langle W(C_{\rm circle}) \rangle =\frac{2e^{-{g^2 (1+N)\over 8N}}}{ N!\ g^{2N}  }
\int_0^\infty dt\ e^{-t} B(t g^2)\ ,
\ee
where
\be
B(x)\equiv \  x^{N-{1\over 2}} I_1(\sqrt{x})\ .
\ee
Consider the perturbation series of $\langle W\rangle$.
Using the power series expansion of Bessel function $I_1$, we find
\bea
\langle W(C_{\rm circle}) \rangle &=&\frac{e^{-{g^2(1+N)\over 8N}}}{N! }\sum_{k=0}^\infty
\frac{(N+k)!g^{2k}}{4^k k!(k+1)!}
\nonumber\\
&=&\frac{1}{N! }\sum_{n=0}^\infty c_n \ g^{2n}\ ,\qquad c_n=\frac{(-1)^{n}}{8^n }\sum_{k=0}^n \frac{(-1)^{k}2^k(N+k)! (1+{1\over N})^ {n-k}}{ k!(n-k)! (k+1)!}\ .
\nonumber
\eea
{}For large $n$, $c_n\sim b^n/n!, \ b=(N-1)/8N$.
Thus the perturbative series for this observable 
has an infinite radius of convergence.
As we will see this is not the case for gauge theories with less supersymmetry,
where the perturbation series will be divergent.

Consider now the planar $N\to\infty $ limit, with $\lambda= g^2N$ fixed. Using
 $(N+k)!/N! \approx N^k$,  we see that $\langle W(C_{\rm circle}) \rangle \to 2I_1(\sqrt{\lambda})/\sqrt{\lambda}$,
which shows that  the series expansion in planar diagrams also  has an infinite radius of convergence for this observable.

In this paper we  consider the following
theories: the  ABJM theory \cite{Aharony:2008ug}, the ${\cal N}=2$ superconformal $SU(2)$ Yang-Mills theory with four fundamental
hypermultiplets, and the  ${\cal N}=2^*$  $SU(N)$   theory obtained by adding a mass term for the adjoint hypermultiplet of ${\cal N}=4$ 
super Yang Mills (SYM) theory.  Various properties of these ${\cal N}=2$ SYM theories are discussed in \cite{Seiberg:1994aj}.

In section 2 we consider the exact expression for the partition function of $U(2)\times U(2)$ ABJM theory on $S^3$ obtained by Kapustin et al in \cite{Kapustin:2009kz}. We determine the asymptotic behavior of the perturbation series and show that it is summable in the sense of Borel.
In section 3 we show that the partition function in the zero-instanton sector of  the ${\cal N}=2$ superconformal $SU(2)$ Yang-Mills theory on $S^4$
can be written as a Borel transform. This formula explicitly exhibits the singularity structure and the convergence radius of the associated series.
We also determine the asymptotic behavior of the perturbation series by a saddle point evaluation and discuss the one-instanton sector. 
The analysis is then extended
to the expectation value of a circular Wilson loop operator. 
Section 4 contains several calculations on the  ${\cal N}=2^*$  SYM theory. We first consider the $SU(N)$ theory in the planar
limit and compute the distribution of eigenvalues and the expectation value of the circular Wilson loop in different regimes of the gauge coupling and mass parameters.
We then consider the $SU(2)$ gauge theory. We discuss different properties  for the partition function, Wilson loop
and 't Hooft loop operators, we comment on  the structure and  asymptotic behavior of the perturbative series,
the role of instantons and on the small and large mass behavior. Section 5 contains  concluding remarks.


\section{ABJM matrix model}

\noindent 1. The ABJM theory \cite{Aharony:2008ug} is constructed by using two copies of Chern-Simons theory with gauge group $U(N)\times U(N)$
and opposite  levels $k, \ -k$.
The general formula for the partition function of ABJM theory on $S^3$ is given by \cite{Kapustin:2009kz}
\be 
Z_{\rm ABJM} = {1\over 2^{2N}N!^2} \int {d^N\mu\over (2\pi )^N} {d^N\nu\over (2\pi )^N} 
\frac{ \prod_{i<j} \sinh^2 \left({\mu_i-\mu_j \over 2}\right)
\sinh^2 \left({\nu_i-\nu_j \over 2}\right)}{
\prod_{i,j} \cosh^2 \left({\mu_i-\nu_j \over 2}\right)}
\ e^{{ik\over 4\pi}\sum_{i=1}^N \big(\mu_i^2-\nu_i^2\big)}\ .
\label{zast}
\ee
$1/k$ plays the role of coupling constant. 
By some simple manipulations the partition function can be casted in the 
following form \cite{Kapustin:2010xq,Marino:2011eh}
\be 
Z_{\rm ABJM}= {1\over 2^{2N} N!} \sum_{\sigma\in S_N} (-1)^{\varepsilon(\sigma)}
 \int {d^N x\over (2\pi k)^N} 
\frac{1}{\prod_i \cosh({x_i\over 2}) \cosh \left({x_i-x_{\sigma(i)} \over 2k}\right)
}\ .
\ee
Different aspects of the partition function have also been discussed in \cite{Drukker:2010nc,Drukker:2011zy}.

\bigskip

We begin the discussion with a case where calculations can be made explicitly. The ABJM model with gauge group $U(2)\times U(2)$, i.e. $N=2$.
In this case we have
\bea
Z_{\rm ABJM}(N=2) &=& {1\over 32} \int {d x_1 dx_2\over (2\pi k)^2} \frac{1}{\cosh({x_1\over 2})\cosh({x_2\over 2})}
 \left(1  -   \frac{1}{\cosh^2 \left({x_1-x_2 \over 2k} \right)}\right)
\nonumber\\
&=& {1\over 32 k^2}  \left(1 -  J(k) \right)\ ,
\eea
where
\be
J(k)\equiv  {1\over 4\pi^2} \int {d x_1 dx_2} \frac{1}{\cosh({x_1\over 2})\cosh({x_2\over 2})\cosh^2 \left({x_1-x_2 \over 2k} \right)} \ .
\ee
Using the Fourier integrals:
\bea
&&\frac{1}{\cosh^2  \left({x_1-x_2 \over 2k} \right)} ={2k^2}\int dt\ {t\over \sinh({\pi kt})}e^{it (x_1-x_2)}
\nonumber\\
&&\int dx \frac{1}{\cosh {x\over 2} }  e^{it x}=   {2\pi\over \cosh(\pi t )}
\eea
we can  perform the integrals over $x_1, x_2$. We get
\be
J(k)={2\over\pi^2} \int dv {1\over \cosh^2({v\over k}) } {v\over \sinh({v})}\ ,\qquad v=\pi t k\ .
\label{zasj}
\ee
This integral can be evaluated by residues, as done in \cite{Okuyama:2011su}. The result is a finite sum of $k$ different contributions of
$k$-depending terms. For the purpose of studying the power  series expansion in $1/k$, it is more convenient to work with the integral
representation (\ref{zasj}) of $J(k)$.
We introduce
\be
I(k)\equiv - {\pi^2\over 2}\int {dk\over k^2} \ J(k) =\int dv \ {\tanh({v \over k})\over \sinh({v})}\ .
\ee
Using the formulas
\be
\tanh({v\over k}) =\sum_{m=1}^\infty 4^m \left(4^m-1\right) \ \frac{B_{2 m} }{(2 m)!}
\left(\frac{v}{k}\right)^{2 m-1}\ ,
\ee
\be
\int_{-\infty}^\infty dv\ {v^{2m-1}\over \sinh(v)} = (-1)^{m+1} {1\over m} \ \pi^{2m}(4^{m}-1)B_{2m}\ ,
\ee
where $B_{2n}$ are the Bernoulli numbers,
we find
\be
I(k)= -k \sum_{m=1}^\infty (-1)^m {\left(B_{2m}\right)^2\over m(2m)!} (4^{m}-1)^2  \left( { 2\pi\over k}\right)^{2m}\ .
\ee
Hence
\be
J(k) =-{2k^2\over\pi^2} I'(k) = -{2k^2 \over  \pi ^2}\sum_{m=1}^\infty
(-1)^m  (2 m-1) \left(4^m-1\right)^2   {\left(B_{2m}\right)^2\over m(2m)!} \left( { 2\pi\over k}\right)^{2m}\ ,
\ee
and
\bea
Z_{\rm ABJM}(N=2,k)
&=& {2 \over ( 4\pi )^2}\sum_{m=2}^\infty
(-1)^m  {(2 m-1)\over 2m} \left(4^m-1\right)^2   {\left(B_{2m}\right)^2\over (2m)!} \left( { 2\pi\over k}\right)^{2m}
\nonumber\\
&=& {\pi^2\over 64k^4} \left(1-\frac{4 \pi ^2}{3 k^2}+\frac{289 \pi ^4}{90 k^4} -\frac{3844 \pi ^6}{315 k^6}+...\right)\ .
\label{zart}
\eea
It is interesting to examine  the asymptotic behavior of this series.
We use the familiar  formula:
\bea
\label{berno}
\frac{B_{2m}}{(2m)!} &=& (-1)^{m+1}\frac{2}{(2\pi)^{2m}}\zeta(2m) 
\nonumber\\
 &=& (-1)^{m+1}\frac{2}{(2\pi)^{2m}}\big(1+ 2^{-2m}+3^{-2m}+4^{-2m}+...\big)\ ,
\eea
We find the following behavior:
\be
Z_{\rm ABJM}(N=2) \sim 
\sum_{m\gg 1} (-1)^m (2m)!  \left({ 2\over \pi k}\right)^{2m}\ .
\ee
Thus the coefficient of $1/k^n$ increases as $n!$.
The series (\ref{zart}) is therefore an asymptotic series in the sense of Poincar\' e. 
As more terms of the series are included, the sum approaches the true value of $Z_{\rm ABJM}(N=2,k)$
until  it begins to deviate leading to a divergent asymptotic behavior.
Minimizing a generic term with respect to $m$ one finds the optimal truncation (see discussion at the end of section 3).

The series is divergent and at this point the main question is whether it is Borel summable.
The associated Borel series can be defined in the standard way: one introduces a
 ``1", 
\be
1= {1\over (2m)!} \int_0^\infty dt \ t^{2m}e^{-t} \ ,
\ee
and write 
\be
Z_{\rm ABJM}(N=2,k)
= {2 \over ( 4\pi )^2}\sum_{n=2}^\infty \int dt\ e^{-t}\ 
(-1)^m  {(2 m-1)\over 2m} \left(4^m-1\right)^2   {\left(B_{2m}\right)^2\over (2m)!^2} \left( { 2\pi t\over k}\right)^{2m}\ .
\ee
Because the series is divergent, commuting integral and sum does not give an equivalent result.
Instead it defines the associated Borel series $B(t)$,
\be
\hat Z_{\rm ABJM}(N=2,k)
\equiv \int_0^\infty dt\ e^{-t}\ B(t/k)\ ,
\label{far}
\ee
\be
B(t)\equiv{2 \over ( 4\pi )^2}\sum_{n=2}^\infty 
(-1)^m  {(2 m-1)\over 2m} \left(4^m-1\right)^2   {\left(B_{2m}\right)^2\over (2m)!^2} \left(  2\pi t\right)^{2m} \ .
\ee
The perturbation series will be Borel summable if the integral (\ref{far}) can be done, which requires
that possible singularities in $B(t)$ do not lie on the positive real $t$ axes (or, if they do, they are integrable, like e.g. in the case of a Dirac delta function singularity).
Asymptotically, $B(t)$ behaves as follows
\be
B(t) \sim \sum_{m\gg 1} (-1)^m   \left(  {2 t\over\pi } \right)^{2m}\ ,
\ee
which has a finite radius of convergence. More precisely, inserting the full expansion (\ref{berno}),
we find that $B(t)$ is a sum of series with finite radii of convergence, the minimum radius being at
$R_*={\pi\over 2}$ so that all series are convergent for  $|t|<R_*$. Other series have radius of convergence of the form $R={p\pi\over 2}$, where $p$ is a positive integer.
There are singularities in $B(t)$  on the imaginary axes at $t=\pm {i\pi\over 2}$ and more generally  
at $t= {i\pi p\over 2}$ where $p \in {\bf Z}$.

Therefore we have shown that the perturbation series for the $U(2)\times U(2)$ ABJM partition function is Borel summable.
This is an expected result, because there are no instantons in three dimensions.
We recall that a singularity of $B(t/k)$ on the real positive axes of $t$ at $t/k=a $
would have implied the presence of euclidean solutions with 
finite action  $e^{-t}\sim e^{-a k}$.


\bigskip 

\noindent 2. We now consider an alternative treatment. 
In the present case, having closed analytic expressions, there is actually no need of artificially introducing a Borel associated sum by
a Laplace transform. Indeed, one may define
a slight modification of the Borel transform that works equally well and is more convenient in  cases where
the one-loop determinants involve hyperbolic trigonometric  functions.
Given an asymptotic  series 
\be 
F(g)=\sum_{n=0}^\infty f_n\ g^n\ ,
\ee
one can introduce an  associated series ${\cal S}(z)$ defined by
\be
F(g) =2 \int_0^\infty dz\ {z\over \sinh z} {\cal S}(gz)\ ,
\label{tru}
\ee
with the expansion
\be
 {\cal S}(x)= \sum_{n=0}^\infty s_n\ x^{n}\ ,
\ee
where $s_n$ is related to the original series coefficient $f_n$ by the formula
\be
 f_n= 2^{-n} \left(2^{n+2}-1\right) \zeta (n+2) (n+1)! \ s_n\ .
\ee
Here we have used the  formula
\be
2\int_0^\infty dv\ {v^{n+1} \over \sinh(v)} = 2^{-n} \left(2^{n+2}-1\right) \zeta (n+2) (n+1)!\ .
\ee
Asymptotically, $\zeta(n)\to 1$, therefore
\be
 f_n\sim   (n+1)!\ s_n\ ,
\ee
like in the Borel case. 
Thus the transformation (\ref{tru}) has essentially the same effect of the standard Borel transform.
In the present $U(2)\times U(2)$ case, 
we do not know how to express $Z_{\rm ABJM}$ as a Borel transform in terms of closed analytic formulas, but we
can express $Z_{\rm ABJM}$  in terms of the above ${\cal S}$-transform  (\ref{tru}). This has some obvious advantages.
In particular, by  (\ref{tru}), one   readily spots an associated series 
with finite radius of convergence, whose expression is determined by a closed analytic formula:
\be
Z_{\rm ABJM}(N=2) =  {1\over 8 \pi^2 k^2}  \int_0^\infty dv  {v\over \sinh({v})}\ {\cal S}({v\over k})  \ ,
\label{jorg}
\ee
with
\be
{\cal S}(x)\equiv \tanh^2(x) \ .
\ee
By means of the  sole structure of ${\cal S}$ one determines that: 
\smallskip

\noindent a)  singularities of ${\cal S}$
arise at complex values for $v$, $v=i k (2n+1){\pi\over 2}$ with integer $n$. 

\smallskip

\noindent b)  the series expansion  of ${\cal S}$ in powers of $1/k$ has a finite radius of convergence ${1\over |k|}<{\pi\over 2}$.

\smallskip

\noindent c)  the perturbation series for the partition function has an asymptotic behavior $(2n)!$ for a term $k^{-2n}$.

\smallskip

\noindent d)  the perturbation series is summable as the poles of  the integrand do not occur at the positive real axes.

\bigskip

\noindent 3.  Consider now the general $U(N)\times U(N)$ case. 
A way to understand the expected asymptotic behavior of the perturbative series is by  a closer look at the partition function (\ref{zast}). For this,
it is convenient to write down the hyperbolic cosines in the familiar infinite product representation,
\be
\cosh x=\prod_{n=1}^\infty \left(1+ \frac{4x^2}{\pi^2(2n-1)^2}\right)\ .
\ee
The perturbation series arises in expanding the integrand (except the exponential factor)
in powers series of $\mu_i,\ \nu_i$ and then integrating over $\mu_i,\ \nu_i$
(again this involves an illegitimate  commutation of sums and integrals).
The series for the integrand have finite radii of convergence determined by the singularities  at
 $\mu_i=\nu_j+i (2l+1)\pi$ where the cosh factors in the denominator vanish.
Each integral over $d\mu_i $ or $d\nu_i$ of given factors $\mu_i^{2n_i}$,  $\nu_i^{2m_i}$ times the exponential factor in (\ref{zast}) introduce
factorials $n_i!$, $m_i!$ which will lead to a divergent behavior of the perturbative series.
The fact that the final divergent perturbative series is summable is obvious because it originates from the integral in (\ref{zast}), which  is well defined and convergent, in particular, it does not hit any of
the singularities in the complex plane.
A  direct way to determine
the asymptotic behavior of the perturbative series is
by means of saddle-point techniques. Defining $\kappa=1/k^2$, the coefficient $d_n$ of the $n$-th term $\kappa^n$ of the power series in $\kappa $ is obtained by
\be
d_n ={1\over 2\pi i} \oint {d\kappa\over \kappa ^{n+1}}\ Z_{\rm ABJM} (N,\kappa) \ ,
\ee 
where the contour surrounds the point $\kappa=0$.
For large $n$, the integrals over $\kappa $ and over
$\mu_i,\nu_i$ are dominated by saddle points in the complex plane, e.g. $\mu_i=\nu_j+i (2l+1)\pi+O(1/n)$, etc.
We will employ this method in discussing the perturbative series for  ${\cal N}=2$ super Yang-Mills theories.

\section{${\cal N}=2$ $SU(2)$ theory with $N_f=4$ massless hypermultiplets} 

Consider first ${\cal N}=2$ $SU(N)$ theory with ${\cal N}_f=2N$ fundamental massless flavors. This theory 
has  superconformal symmetry. 
The partition function on $S^4$ is given by \cite{Pestun:2007rz}
\be
Z_{\rm SCF} = c_N g^{-(N^2-1)}   \int d^{N-1}a \prod_{i<j} (a_i-a_j)^2\ e^{-{8\pi^2\over g^2}\sum_i a_i^2}
Z_{\rm 1-loop}(a)\  |Z_{\rm inst}(a,g^2)|^2\ ,
\ee
where
\be
Z_{\rm 1-loop}(a) = \frac{\prod_{i<j} H^2(a_i-a_j)} { \prod_{i}H^{2N}(a_i)}\ ,\qquad H(x)\equiv \prod_{n=1}^\infty 
\left(1+{x^2\over n^2}\right)^n e^{-{x^2\over n}}\ ,
\ee
and $\sum_i a_i=0$.
$c_N$ is a numerical constant independent of $g$. 
Following \cite{Passerini:2011fe}, we introduced the function $H$  related to the Barnes $G$ function by the relation
\be
H(x)=e^{-(1+\gamma )x^2}G(1+ix)G(1-ix)\ ,
\ee
where $\gamma $ is the Euler's constant.
The instanton  factor has a more complicated form and general formulas are given in \cite{Nekrasov:2002qd,Nekrasov:2003rj,Alday:2009aq}.
For the case of $S^4$, one must  fix the Nekrasov deformation parameters   \cite{Nekrasov:2002qd} to $\varepsilon_1 =\varepsilon_2=1/r$,
where $r$ is the radius of $S^4$ \cite{Okuda:2010ke}. In what follows we set $r=1$.

Various properties of the partition function and circular Wilson loop for the 
large $N$  $SU(N)$ theory have  been recently discussed in \cite{Rey:2010ry,Passerini:2011fe}.
In this section we focus on the $SU(2)$ theory. 
For an  $SU(2)$ gauge group, the  partition function reduces to
\be
Z_{\rm SCF} ^{SU(2)}= 128\pi^{5/2} \ g^{-3} \int_{-\infty}^\infty da \ a^2  e^{-{16\pi^2\over g^2}\ a^2}
\prod_{n=1}^\infty \frac{
(1+{4a^2\over n^2})^{2n} }{(1+{a^2\over n^2})^{8n}}
\ |Z_{\rm inst}^{SU(2)}|^2\ ,
\label{vamos}
\ee
where we have fixed the numerical constant $c_N$ to have $Z_{\rm SCF}^{SU(2)}=1$ at $g=0$.
Now
\be
\prod_{n=1}^\infty \frac{
(1+{4a^2\over n^2})^{2n} }{(1+{a^2\over n^2})^{8n}} ={H^2(2a)\over H^8(a)}=\frac{ |G(1+i2a)|^4}{  |G(1+ia)|^{16}}\ .
\ee

Consider  the ($k=0$)-instanton sector, obtained by setting $|Z_{\rm inst}^{SU(2)}|\to 1$ 
in (\ref{vamos}). By the change of variable
\be
a^2= \alpha \ t  \ ,\qquad \alpha \equiv \frac{g^2}{16\pi^2}\ ,
\ee
we get 
\be
Z_{\rm SCF}^{SU(2)}\bigg|_{k=0} = \frac{2}{\sqrt{\pi\a }}\ \int_0^\infty dt\ e^{-t} \ B(\a t )\ ,
\ee
where
\be
B(x)= \sqrt{x}\ \prod_{n=1}^\infty \frac{
(1+{4x\over n^2})^{2n} }{(1+{x\over n^2})^{8n}}\ .
\ee
Remarkably, in the zero instanton sector the partition function is already in the form of a Laplace transform of a Borel associated function.

This compact form of $B(x)$ explicitly exhibits the singularities in the Borel plane. There are singularities
at 
\be
\alpha t_s=  -n^2\ ,\ \ \ \ \ n=1,2,...
\ee
The expansion in powers of $\a t$ of $B(\a t)$ has a finite radius of convergence
given by $|\alpha t_*|=1$.   Integrating over $t$ adds a factor $n!$ to each $n$-th term $\a^n$ of this expansion.
Since the series for $B(\a t)$ has a finite radius of convergence, the $n$-th term of the perturbative series
will  have the asymptotic behavior $ n!$ (more precisely, $(-1)^n n!$ since the singularity appears at a negative $t$).

It is easy to compute the first few terms of the perturbative expansion. Using
\be
B(x) =  \sqrt{x}\ \exp\left[ -2 \sum_{k=2}^\infty {(-1)^k\over k}\ x^k \zeta(2k-1)\ (2^{2k}-4)\right]\ ,
\ee
we find 
\bea
Z_{\rm SCF}^{SU(2)}\bigg|_{k=0} &=&
1-45 \alpha ^2 \zeta (3)+525 \alpha ^3 \zeta (5)+\frac{8505}{8} \alpha ^4 \left(4
   \zeta (3)^2-7 \zeta (7)\right)
\nonumber\\
&-& \frac{31185}{4} \alpha ^5 (20 \zeta (3) \zeta
   (5)-17 \zeta (9))
\label{car}\\
&-&
\frac{135135}{16} \alpha ^6 \left(72 \zeta (3)^3-200 \zeta
   (5)^2-378 \zeta (3) \zeta (7)+341 \zeta (11)\right)+O\left(\alpha ^7\right)\ .
\nonumber
\eea
Note that no $\pi $ factors appear in the perturbative expansion; it is  an expansion in powers of $\alpha =g^2/16\pi^2$ and
the associated Borel sum is convergent for $|\alpha t| <1 $.
The partition function is well defined for any $\a $ as there are no singularities in the integration region.

The divergent perturbative series (\ref{car}) is what one would have obtained if instead of using matrix model localization techniques the partition function had been computed
using Feynman diagrams.\footnote{For example, the  three-loop term with $\zeta(3)$ coefficient
in the similar perturbative series for the Wilson loop, see (3.26), has been reproduced by computing three loop Feynman diagrams  in \cite{Andree:2010na}.}
 In the present context, it is clear that
the divergence of the perturbative series emerges artificially, in the process of commuting the integral over $t$ with the sum defining the expansion of $B$ in powers of $\a t$.
There is no divergence in the original integral defining the partition function.
The singularities of $B(\a t)$ lie on the negative real axes, which means that the integral over $t$ can be performed and therefore the divergent perturbative series computed by Feynman diagrams is summable
in the Borel sense.
It is the Feynman diagrammatic method which introduces the divergent behavior, because computing Feynman diagrams 
requires commuting the functional integral with the expansion in series of the exponential of the interaction terms.
The convergent Borel associated sum is indeed the true result of the theory, as defined by the functional integration; the divergent perturbative series 
here (and in the Feynman diagram calculation) arises artificially due to a mathematically illegitimate manipulation.

\medskip

 Let us now compute the asymptotic behavior of the perturbative series  by a saddle point calculation at large $n$.
Writing
\be
Z_{\rm SCF}^{SU(2)}\bigg|_{k=0} =\sum_{n=0}^\infty d_n\ \a^n\ ,
\ee
we have
\be
d_n ={1\over 2\pi i} \oint \frac{d\a}{\a^{n+1}} Z_{\rm SYM}^{SU(2)}\bigg|_{k=0} = {1\over 2\pi i} \frac{2}{\sqrt{\pi\a }} \oint d\a \int_0^\infty  dt\ e^{-S}\ ,
\ee
\be
S(\a,t) = t + (n+1)\log \a -\log B(\a t)\ .
\label{tras}
\ee
For large $n$, the saddle point equations in $\a $ and $t$ are
\be
{n\over \a} = \frac{t B'(\a t)}{B(\a t)} \ ,\qquad 1 = \frac{\a B'(\a t)}{B(\a t)}\ ,
\ee
i.e. $t_*=n$ and $\a $ satisfying the equation $\a B'(\a n)/B(\a n)=1$.
This equation has an infinite number of solutions, with $\a n$ near the zeroes and singularities of $B(\a n)$.
The dominant saddle point (with smallest $S$ and large eigenvalues of the Jacobian matrix), 
appears near the first pole, $\a_* n= -1+O(1/n)$.
Computing $S$ at this point, we find
\be
d_n \sim e^{-S(\a_*,t_*)} \sim (-1)^n  n^n e^{-n} \sim (-1)^n\ n!\ ,
\ee
as anticipated above.

\bigskip
It is well known that the existence of instantons can lead to singularities in the Borel transform.
Having closed analytic expressions, it is therefore interesting to  study how  
 the perturbation series behave in the non-zero instanton sectors and to determine the extent to which  possible singularities
produced by instantons could affect the convergence of integrals.
 Consider for  concreteness the one-instanton sector. The one-instanton contribution is given by
\bea
Z_{\rm one-inst} ^{SU(2)} &=& 4 e^{-{8\pi^2\over g^2}} \int {dy\over 2\pi} 
\left( \frac{y^{4}}{ \big( (y-a)^2+1\big)\big( (y+a)^2+1\big)} -1\right)
\\
&=&  e^{-{8\pi^2\over g^2}}\  (a^2-3)\ .
\eea
Thus
\be
Z_{\rm SCF}^{SU(2)}\bigg|_{k=1} = \ \int_{-\infty}^\infty da\ a^2 e^{-{(a^2+1)\over \a} } (a^2-3)^2\  \frac{H^2(2a)}{H^8(a)}
\ .
\ee
This integral is well defined and convergent.
Now the partition function is no longer in the form of a Laplace transform of a Borel function, because of the presence of the instanton weight
 $e^{-{|k| \over\a }}$. 
To write it as a Borel transform one would need to put it in the form
\be
Z_{\rm SCF}^{SU(2)}\bigg|_{k=1} = \int_0^\infty dz\ e^{-{z\over\a}}  \ F( z )\ .
\ee 
This can be done by a change of coordinate, $a^2+1=z$. We find
\be
F(z)\equiv \theta(z-1) \sqrt{z-1}\ (z-4)^2\ f(z-1)\ .
\ee
This has a branch cut singularity at $z=1$ (more generally, $z=|k|$ in the $|k|$ instanton sector), but obviously it does   not affect the convergence of the integral.
Let us now see how this singularity is manifested in the saddle-point evaluation. The analysis is as follows.
In the $1$-instanton sector, instead of (\ref{tras}) we would have
\be
S(\a,a) = {a^2\over\a } + n\log \a -2\log \frac{H(2a)}{H^4(a)} +{1 \over\a } -  \log a^2 -\log(a^2-3)^2 \ .
\label{atras}
\ee
The saddle points which lie near the singularities of $\log H(2a)/H^4(a) $ are not significantly modified.
The main difference is that now  a new saddle point emerges, located at
$a^2_*= \a _* =1/n $. Since  $a^2\sim 0$, terms
$\log H(2a)/H^4(a) $ and $\log(a^2-3)^2$ are irrelevant. This is indeed the dominant saddle point.
The corresponding saddle point in the $|k|$ instanton sector is located at $a^2_*= \a _* =|k|/n $.
One finds
\be
d_n \sim {n^n\over |k|^n}\ e^{-n}\sim n!\ |k|^{-n}\ ,
\ee
which shows that the Borel function must have a singularity at $z=|k|$. This is precisely the location of the branch-cut singularity.

In conclusion, the integral defining the partition function is convergent in the one-instanton sector.
The instanton produces the expected singularity in the Borel function, which  in the present case turns out to be harmless.


\subsubsection*{Wilson loop}

Similar considerations can be made for physical observables such as expectation values of Wilson loop operators.
Do they exhibit the same asymptotic behavior?
We restrict to the zero-instanton sector. For a circular Wilson loop located at the equator of $S^4$ in the spin-$j$ representation,
one has \cite{Pestun:2007rz}
\be
\langle {\rm tr}_j {\rm Pexp}\big(\int Adx+i\Phi_0 ds\big)\rangle =Z^{-1} \int_{-\infty}^\infty
da \ a^2  e^{-{16\pi^2\over g^2}\ a^2}
\prod_{n=1}^\infty \frac{
(1+{4a^2\over n^2})^{2n} }{(1+{a^2\over n^2})^{8n}} \left(\sum_{\ell =-j}^j e^{4\pi \ell a}\right)\ .
\ee
Consider the  term $e^{4\pi \ell a}$ ($ \ell$ can be integer or half-integer).
We have
\be
\langle e^{4\pi \ell a}\rangle =\frac{\int_0^\infty dt\ e^{-t} \ B(\a t )\cosh(4\pi \ell \sqrt{\a t})}{\int_0^\infty dt\ e^{-t} \ B(\a t )}\ .
\ee
This can be expanded in powers of $\alpha $ as done in \cite{Pestun:2007rz}. We find
\bea
\langle e^{4\pi \ell a}\rangle &=& 1+12 \pi ^2 \alpha  \ell^2+40 \pi ^4 \alpha ^2 \ell^4+\alpha ^3 \left(\frac{224 \pi ^6
   \ell^6}{3}-720 \pi ^2 l^2 \zeta (3)\right)
\nonumber\\
&+&\alpha ^4 \left(96 \pi ^8 \ell^8-5760 \pi ^4
   \ell^4 \zeta (3)+12600 \pi ^2 \ell^2 \zeta (5)\right)+O(\a^5)
\eea
This expression has been checked up to the three-loop  $\a^3$ order in \cite{Andree:2010na} by explicit evaluation of the Feynman diagrams  in perturbation theory.

{}A term with a given power $p$ of $\alpha $ is multiplied by a polynomial in $\ell $ of degree $2p$.
Terms with lower power of $\ell $ have coefficients which increase more rapidly with $p$.
This is because each power $\ell^{2s } t^{s }$ is accompanied by a factor $1/(2s )!$ coming from the power expansion of $\cosh(4\pi \ell \sqrt{\a t})$.
Thus the terms in the perturbative series for the Wilson loop that give rise to the most divergent behavior of the series are those with lower power of
$\ell $ in the polynomial of degree $2p$ multiplying $\a^p$. For them, $s =1$, so such terms clearly present the same asymptotic behavior  $(-1)^n n!$
that appeared in $Z_{\rm SCF}^{SU(2)}$.

A different behavior can  be found in a limit  where $\ell \to\infty $ and $\a\to 0$ with $\sqrt{\a } \ell $ fixed.
This gives
\bea
\langle e^{4\pi \ell a}\rangle &\rightarrow &
\frac{\int_0^\infty dt\ e^{-t} \ \sqrt{\a t }\cosh(4\pi \ell \sqrt{\a t})}{\int_0^\infty dt\ e^{-t} \ \sqrt{\a t }}
\nonumber\\
&=& e^{4\pi ^2 \alpha  \ell^2} \left(1+8 \pi ^2 \alpha  \ell^2\right) 
\nonumber\\
&=& e^{ g^2  \ell^2\over 4} \left(1+\frac{g^2  \ell^2}{2}\right) \ .
\label{adru}
\eea
In this limit
the one-loop determinant was set to 1, $B(\a t)\to \sqrt{\a t}$. The result   must then be the same as in ${\cal N}=4$ SYM as described in the introduction.
Indeed, by explicitly computing  the integral (\ref{druk}) with $N=2$, we exactly reproduce (\ref{adru}) with $\ell =1/2$.
Now the expansion in powers of $\alpha $  has an infinite radius of convergence and $\langle e^{4\pi \ell a}\rangle $ is an exponentially increasing function of the parameter $\alpha \ell^2$
which starts at 1. 
 
\medskip

An interesting question regards the  best resolution that the asymptotic perturbative series defining
the Wilson loop can provide.
This is done in the standard way by considering the partial sum 
\be
W_{n_0}=\sum_{n=1}^{n_0} c_n \a^n\ ,\ \ \ \ \ c_n\sim (-1)^n n! R^{-n}\ .
\ee
The optimal truncation is then obtained by using the Stirling approximation and minimizing $|c_n|\sim n! R^{-n}$ with respect to $n$.
One finds $n_0=R/\a $, for which the best resolution (or ``non-perturbative ambiguity") is $\epsilon =\sqrt{2\pi R/\a}\ e^{-R/\a}$, where we used the Stirling formula.
In the present case, we have closed expressions which allow an accurate determination of $R$, which is nothing
but the radius of convergence of the associated Borel series.
This gives $R=1$, so that $n_0={16\pi^2/g^2}$ and the best  resolution is $\epsilon=\sqrt{2\pi /\a}\ e^{-1/\a}\sim  e^{-16\pi^2/g^2}$.
For example, if $g=1$, then $\a \sim 0.006 $; this gives $n_0=158$ and $\epsilon=10^{-67}$.
But if $g=8$, $\a \sim 0.4$, then $n_0\sim 2.4$ and the asymptotic formulas are not applicable.
It should be noted that here the ``non-perturbative ambiguity" $\epsilon =\sqrt{2\pi R/\a}\ e^{-R/\a}$  is  unrelated
to  instanton contributions that would provide a ``completion" of the series,  since the origin of the divergence in the asymptotic series are singularities in the one-loop determinant.
In particular, in the ABJM model discussed in section 2 there is a similar non-perturbative ambiguity $\epsilon=O(e^{-{\pi k\over 2}})$ despite the fact that there are no instantons.



\section{  ${\cal N}=2^*$ $SU(N)$   Super Yang-Mills theory}

It is of interest to consider the behavior of the perturbative series and of loop operators in a case where there is no superconformal invariance.
Typically, in such cases the one-loop determinant contains a divergence which needs to be renormalized \cite{Pestun:2007rz}.
One case where this divergence is innocuous is  in the ${\cal N}=2^*$ $SU(N)$ SYM,
obtained by deformation of ${\cal N}=4$ by adding a mass $M$ for the adjoint hypermultiplet.
In this case the divergence is a numerical factor  (in particular, it has no incidence on the expectation value of physical observables). We remove this divergent numerical factor from the partition function by inserting $e^{-{2M^2\over n}}$ in the infinite product over $n$ in the integrand. The  partition function on $S^4$ reads\cite{Pestun:2007rz}
\be
Z^{N=2^*} =  \int d^{N-1}a \ \prod_{i<j} (a_i-a_j)^2  e^{-{8\pi^2\over g^2}\sum_i a_i^2}
Z_{\rm 1-loop}(a,M)\ |Z_{\rm inst}^{SU(N)}|^2\ ,
\ee
with 
\bea
Z_{\rm 1-loop}(a,M) &=& \prod_{i<j}  \frac{H^2(a_i-a_j)} { H^2(a_i-a_j-M) }
\nonumber\\
&=& \prod_{i<j}\prod_{n=1}^\infty \frac{  (1+{(a_i-a_j)^2\over n^2})^{2n} }{ \left(1+{(a_i-a_j-M)^2\over n^2}\right)^{n}\left(1+{(a_i-a_j+M)^2\over n^2}\right)^{n}e^{-{2M^2\over n}}}\ .
\eea
${\cal N}=2^*$ $SU(N)$ super Yang-Mills theory interpolates between  ${\cal N}=4$ $SU(N)$ SYM  and pure ${\cal N}=2$ $SU(N)$ SYM 
as the mass parameter is varied from zero to infinity. The interpolation seems to be smooth (for example, on ${\bf R}^4$ the Seiberg-Witten curve presents no discontinuous or singular behavior for any $M$ \cite{Seiberg:1994aj}).
For large $M$, the ${\cal N}=2^*$ theory may be viewed 
as a regulated version of pure ${\cal N}=2$ theory where one has an UV cutoff determined by the mass.
More generally, given an asymptotically free ${\cal N}=2$ theory, by adding a sufficient number of massive hypermultiplets one gets
a theory that is well defined in the UV and for large masses it provides a regulated version of the ${\cal N}=2$ theory that one wishes to study.
This interpretation was exploited in  \cite{Pestun:2007rz} to recover the  $\beta $ function of ${\cal N}=2$ SYM from the partition function.
This observation  will be reproduced here in the planar limit. In addition, we will also study the small mass behavior  
and loop operators in the $SU(2)$ theory.

\subsection{Large $N$  limit and Wilson loop}

We consider planar limit, i.e. $N\to\infty $ with $\lambda =g^2N$ fixed.
In this limit  $|Z_{\rm inst}^{SU(N)}|\to 1$.
We will compute the integral in the saddle point approximation.
We use the  notation of \cite{Passerini:2011fe}, where the similar calculation was carried out for the ${\cal N}=2$ superconformal theory discussed in the previous section. 
Our approach to solve the system is different: now there is an extra parameter, the mass, in addition to the
't Hooft coupling $\lambda=g^2N$, which will lead us to consider different regimes.

The effective action for the eigenvalues is
\be
S(a)= \sum_i  {8\pi^2\over \lambda} a_i^2 -{1\over N} \sum_{i<j} \left(\log (a_i-a_j)^2 + \log \frac{H^2(a_i-a_j) }{
H(a_i-a_j-M)\ H(a_i-a_j+M)}\right)\ .
\ee
The saddle-point equations are therefore
\be
\frac{8\pi^2 a_i}{\lambda}   -{1\over N} \sum_{i\neq j}\left( \frac{1}{a_i-a_j} - K(a_i-a_j)+{1\over 2}\big( K(a_i-a_j-M)+K(a_i-a_j+M)\big) \right)=0\ ,
\ee
where
\bea
K(x)&=& -{H'(x)\over H(x)}= x \left( \psi(1+i x) +\psi(1-i x)+2\gamma \right)
\nonumber\\
&=& -2\sum_{k=1}^\infty (-1)^k \zeta(2k+1) x^{2k+1}\ ,
\label{opi}
\eea
and $\psi $ is the logarithmic derivative of the $\Gamma $ function. 

Introducing as usual the eigenvalue density
\be
\rho(x)={1\over N}\sum_i \delta(x-a_i)\ ,
\ee
the saddle-point equation can be written as the following integral equation:
\be
\dashint_{-\mu}^\mu dy \rho(y) \left(\frac{1}{x-y} -K(x-y)+{1\over 2}(K(x-y+M)+K(x-y-M))\right)= \frac{8\pi^2}{\lambda}\ x\ .
\ee
Following \cite{Passerini:2011fe}, we do the transform
$$
\dashint_{-\mu}^\mu \frac{dx}{z-x}\ \frac{1}{\sqrt{\mu^2-x^2}}\ \times\ 
$$
The integral equation then takes the form
\bea
\rho(x) &=& \sqrt{\mu^2-x^2}\ \bigg({8\pi\over\lambda } -{1\over 2\pi^2} \dashint \frac{dy}{x-y}\frac{1}{\sqrt{\mu^2-y^2}}
\nonumber\\
&\times &  \int dz\rho(z)\big(2K(y-z)-K(y-z+M)-K(y-z-M)\big) \bigg)\ .
\label{adas}
\eea
We will treat the second term on the RHS of (\ref{adas}) as a perturbation,
which allows us to replace $\rho(z) $ by $n_0\sqrt{\mu^2 - z^2}$ where $n_0=2/(\pi\mu^2)$ is the normalization factor.
The equation can be explicitly solved in two different regimes, $\mu\ll M$ or $\mu\gg M$. 
Since in the perturbative approach $\mu\sim\sqrt{\lambda}$, the first case will lead to a small $\lambda $ expansion, and the second case to a strong $\lambda $ expansion.
 We will now consider these two cases separately.

By rescaling $x, y, z\to  \mu\tilde x, \mu\tilde y, \mu\tilde z$, one gets  new tilde variables which are of order 1.
For $\mu\ll M$ we can expand $K\big(\mu(\tilde y-\tilde z)\pm M)$ in powers of $(y-z)$. One has
\be
2K(y-z)-K(y-z+M)-K(y-z-M)=-2K'(M)\ (y-z) + O\big((y-z)^2\big)\ .
\ee
By computing the integrals we find
\be
\rho(x) \approx \sqrt{\mu^2-x^2}\ \left({8\pi\over\lambda } -{1\over \pi} \ K'(M)\right)\ .
\label{roro}
\ee
The perturbative treatment of the saddle point equation is therefore justified for
\be
\frac{\lambda  K'(M)}{8 \pi ^2}\ll 1\ .
\label{mak}
\ee
The normalization condition  gives
\be
\mu \approx \frac{\sqrt{\lambda }}{2 \pi  \sqrt{1-\frac{\lambda  K'(M)}{8 \pi ^2}}}\ .
\label{mumu}
\ee
The eigenvalue density (\ref{roro}) can be written in the form
\be
\rho(x) \approx {8\pi\over\lambda_{\rm eff} }  \sqrt{\mu^2-x^2}\  ,
\label{roco}
\ee
with
\be
{8\pi\over\lambda_{\rm eff} } \equiv {8\pi\over\lambda } -{1\over \pi} \ K'(M)\ .
\label{runi}
\ee
$K'(M)$ is a monotonically increasing function with the behavior
\be
K'(M)\approx
\begin{cases}
 6\zeta(3) M^2 -10 \zeta(5)M^4\ , &      M\ll 1\ , \\
2\log M e^{1+\gamma} -\frac{1}{6M^2}\ , &      M\gg 1\ , 
\end{cases}
\ee
where we have used (\ref{opi}) and the asymptotic form of the Barnes $G$ function.
\be
G(1+ iM)G(1-iM)\sim \frac{e^{1\over 6}}{A^2} \exp \left(   -(M^2+{1\over 6} )\log M + {3M^2\over 2} +O(M^{-2})\right)\ .
\ee
$A$ is the Glaisher-Kinkelin constant, $A\approx 1.28$.
Therefore, the condition $\mu\ll M$ together with (\ref{mak}) require $\lambda\ll 1$ for any $M$. For large $M$, one has a stronger requirement
$\lambda\ll 4\pi^2/\log M $.

Defining $g_R^2\equiv \lambda_{\rm eff} /N$ and $g_{\rm bare}^2\equiv \lambda /N$,
for large $M$, one has
\be
{4\pi^2\over g_R^2 } = {4\pi^2\over g_{\rm bare}^2 } -{N\over 2} K'(Mr)\approx {4\pi^2\over g_{\rm bare}^2 } -N \log (Mr)\ .
\label{bee}
\ee
where we have restored the dependence on the radius $r$ of $S^4$.
Thus, in the setup  ${\cal N}=2^*$ SYM  is interpreted as ${\cal N}=2$ SYM without matter multiplets with a UV regulator $M$, $g_R$  represents the renormalized gauge coupling.
This interpretation leads to the familiar expression for the one-loop $\beta $ function.
By replacing $Mr$ by $\nu_0/\nu$, $g^2_R(\nu)$  in (\ref{bee}) represents the running gauge coupling of  pure  ${\cal N}=2$ SYM, defined at an energy scale $\nu$. 

\smallskip

Returning to the  ${\cal N}=2^*$ SYM theory for its own sake, we see from  (\ref{mumu})  that 
as $M$ is gradually increased at fixed $\lambda\ll 1 $, the interval $(-\mu, \mu )$ of the eigenvalue distribution will expand. 
The approximation breaks down
long before the denominator becomes singular.

We can now use the above $\rho(x )$ to calculate the expectation value of the Wilson loop operator at weak coupling.
We find
\bea
W(C_{\rm circle})  &=& \int_{-\mu}^\mu dx\ \rho(x)\ e^{2\pi x}
\nonumber\\
 &=& \frac{2I_1\big(\sqrt{\lambda_{\rm eff}(M)}\ \big)}{\sqrt{\lambda_{\rm eff}(M)}}
\nonumber\\
&\approx & 1+\frac{\lambda }{8}+\lambda ^2 \left(\frac{K'(M)}{64 \pi
   ^2}+\frac{1}{192}\right)+O\left(\lambda ^3\right)\ ,
\eea
or simply
\be
W(C_{\rm circle}) \approx  1+\frac{\lambda_{\rm eff}(M) }{8}+\frac{\lambda ^2_{\rm eff}(M)}  {192}+O\left(\lambda ^3_{\rm eff}\right)\ .
\label{wapo}
\ee
Thus in this approximation the complete $M$ dependence in the Wilson loop is in $\lambda_{\rm eff}(M)$.
It would be interesting to reproduce the non-trivial $\lambda^2$ term proportional to $K'(M)$ by a  perturbative gauge theory calculation.
Note that the series (\ref{wapo}) has an infinite radius of convergence, since it arises from the power series expansion of the Bessel function.
The approximation of course breaks down for $\lambda\sim \lambda_0\equiv 8\pi^2/K'(M)$, so convergence is only demonstrated at $\lambda\ll \lambda_0$. 

Using the  above expansions of $K'(M)$, one can explicitly see the small and large $M$ behavior of
the Wilson loop:
\bea
W(C_{\rm circle})\bigg|_{M\ll 1}  & \approx & 1
+\frac{\lambda }{8}+\lambda ^2 \left(\frac{3\zeta(3)M^2}{32 \pi^2}+\frac{1}{192}\right)+O\left(\lambda ^3\right)\ ,
\nonumber\\
W(C_{\rm circle})\bigg|_{M\gg 1}  & \approx & 1+\frac{\lambda }{8}+\lambda ^2 \ \left(\frac{\log(M)}{32 \pi
   ^2}+\frac{\gamma+1}{32 \pi
   ^2}+\frac{1}{192}\right)+O\left(\lambda ^3\right)\ ,
\eea
where the second equation applies as long as $\lambda \log M\ll 1$ due to the condition (\ref{mak}).

\bigskip

Let us now consider the regime $\mu\gg M$.
In this case we can write
\be
2K(y-z)-K(y-z+M)-K(y-z-M)=-\ M^2 \ K''(y-z) + O\big(M^3\big)\ .
\ee
Using the rescaled variables $\tilde y,\tilde z$, it becomes clear that if $\lambda\gg 1$ the integrals will get most of the contributions from the regions
where the argument of $K''$ is large. Therefore we can approximate $K(x)\approx 2x\log x$ so that $K''(x) \approx 2/x$. 
Computing the integrals we now find
\be
\rho(x) \approx \sqrt{\mu^2-x^2}\ {8\pi\over\lambda }\left(1 -M^2\right)\ .
\ee
Therefore the perturbative approach to the saddle point equation is justified for
$ M^2\ll 1$.
{}Now we get
\be
\mu \approx \frac{\sqrt{\lambda }}{2 \pi  \sqrt{1- M^2}}\ .
\label{mulu}
\ee
$\mu\gg M$ requires $\lambda\gg M^2$. 
Thus we have a calculation of the circular Wilson loop in the ${\cal N}=2^*$ theory for $\lambda\gg1 $  at small $M$.
In this regime, the circular Wilson loop in the ${\cal N}=2^*$ theory exhibits the following behavior
\be
W(C_{\rm circle})\approx e^{2\pi\mu} \approx \exp \left[ \sqrt{\lambda }\ \big(1+\frac{  M^2}{2 }\big)\right]\ .
\ee
This shows that the string tension $T$ increases as a small mass parameter is turned on in the ${\cal N}=4$ theory,
 we get $T\approx \frac{\sqrt{\lambda}}{2\pi}\big(1+\frac{  M^2}{2 }\big)$ at $M\ll 1$.

\subsection{${\cal N}=2^*$ $SU(2)$ theory}

A number of interesting properties can be derived explicitly when the gauge group is $SU(2)$. 
For $N=2$ the partition function reads
\be
Z^{{\cal N}=2^*} =  \frac{128\pi^{5/2}}{ g^{3} }
\int_{-\infty}^\infty da \ a^2  e^{-{16\pi^2\over g^2}\ a^2}
\prod_{n=1}^\infty \frac{(1+{4a^2\over n^2})^{2n} }{ \left(1+{(2a-M)^2\over n^2}\right)^{n}\left(1+{(2a+M)^2\over n^2}\right)^{n}e^{-{2M^2\over n}}}
\ |Z_{\rm inst}^{SU(2)}|^2\ .
\ee

Like in the ${\cal N}=2$ superconformal case, we note that in the zero-instanton sector
the partition function can be written as a Laplace transform of the Borel associated function:
we get 
\be
Z^{{\cal N}=2^*} \bigg|_{k=0} = \frac{2}{\sqrt{\pi\a }}\ \int_0^\infty dt\ e^{-t} \ B^{{\cal N}=2^*}(\a t )\ ,
\ee
where
\bea
B^{{\cal N}=2^*}(x) &\equiv &  
\sqrt{x}\ \frac{H^2(2\sqrt{x}) }{H(2\sqrt{x}-M)H(2\sqrt{x}+M)}
\nonumber\\
&=&\sqrt{x}\ \prod_{n=1}^\infty \frac{(1+{4x\over n^2})^{2n} }{\Big(\big(1-i\frac{M}{n}\big)^2+\frac{4 x}{n^2}\Big)^n
   \Big(\big(1+i\frac{M}{n}\big)^2+\frac{4 x}{n^2}\Big)^n e^{-{2M^2\over n}}}\ ,
\eea
and again we have defined
\be
a^2= \alpha \ t  \ ,\qquad \alpha \equiv \frac{g^2}{16\pi^2}\ .
\ee

The associated Borel series defined by the expansion of $B^{{\cal N}=2^*}(x)$ in powers of $x$ has a finite radius of convergence determined by the first singularities
that arise in   $B^{{\cal N}=2^*}(x)$ 
at $x=\a t=- {1\over 4} (1  \pm iM)^2$. 
The  associated Borel series is therefore convergent if 
$$
|x|< {1\over 4}(1+ M^2)\ . 
$$
Also in this case
the perturbative series has asymptotic behavior $n!$, as follows from the fact that the series expansion 
of $B(x)$ around $x =0$ has a finite radius of convergence.
Because the singularities  lie on the complex $t$ plane,  they do not affect the integration over the real positive axes of $t$. Therefore the integration is well defined (the integral is convergent) and the
 divergent perturbative  series  that one would obtain by computing Feynman diagrams is Borel summable. 
We have reproduced  the $n!$  asymptotic behavior  by a saddle-point calculation, similarly as done above in
the ${\cal N}=2$ superconformal case. 

A systematic way to calculate the perturbative series is by writing $B^{{\cal N}=2^*}(x)$ as follows
\be
B^{{\cal N}=2^*}(x)=\sqrt{x}\ \frac{1}{H^2(M)}\ \exp\left[ -\sum_{k=1}^\infty \frac{(-1)^k}{k}
(4x)^k \ p_k(M)\right]\ ,
\label{ares}
\ee
where 
\bea
p_1(M) &=& K'(M)\ ,
\nonumber\\
p_k(M) &=& 2\zeta(2k-1)-\frac{(-1)^k }{  (2k-1)!}\ K^ {(2k-1)}(M) \ ,\quad k>1\ .
\eea
This form is also suitable in order to take the large $M$ limit, that we shall consider in discussing the Wilson and 't Hooft loop 
expectation values.
For the first few terms one finds
\be
Z^{{\cal N}=2^*} = \frac{1}{H^2(M)}\ \left(1+ 6p_1\ \a +30(p_1^2-p_2)\ \a^2+140(p_1^3-3p_1p_2+2p_3)\ \a^3+O(\a^4)\right)\ .
\ee


\medskip

Consider now the effects of instantons.
In particular, the one-instanton contribution is given by \cite{Pestun:2007rz}
\def\td{\tilde }
\def\ep{\varepsilon}
\be
Z_{\rm 1-inst}=  e^{-{8\pi^2\over g^2}}\ \frac{ 2(\tilde M -1)^2\big(\td M^2 +a^2-2 \td M+4\big)}
{ \big( 4+ a^2\big)}\ ,\ \ \ \tilde M= iM+1\ ,
\label{onei}
\ee
where we have set the radius of $S^4$ to 1.
In the case $M=0$ one has $Z_{\rm 1-inst}=0$ and one recovers the result of ${\cal N}=4$ theory.

Just as in the superconformal SYM case, in the one-instanton sector a new saddle point appears in the calculation of  the
coefficient $d_n$ of the $n$th term $d_n\a^n$ of the perturbative series,
but it does not affect the convergence of the integral.
We also see that the prefactor multiplying the instanton weight  does not introduce singularities on the integration region which 
would otherwise represent an obstruction to computing the matrix integral.
Indeed, the contribution (\ref{onei}) has a singularity in the complex plane at $a=\pm 2i$ (and on the negative real axes for the integration variable $t=a^2/\a$).
Therefore it is innocuous. 

\subsubsection{Wilson loop}

Let us now consider the Wilson loop in the zero instanton sector:
\be
\langle e^{2\pi q a}\rangle =   (Z^{{\cal N}=2^*})^{-1} \int_{-\infty}^\infty da \   a^2\ e^{-{a^2\over \a_{\rm eff}(M)}}
\hat B^{{\cal N}=2^*}(a^2,M) \cosh(2\pi q a)\ ,
\label{rizos}
\ee
where we have defined (cf. (\ref{ares}))
\be
\hat B^{{\cal N}=2^*}(a^2,M) =\frac{1}{H^2(M)}\ \exp\left[ -\sum_{k=2}^\infty \frac{(-1)^k}{k}
(2a)^{2k} \ p_k(M)\right]\ ,
\label{hatB}
\ee
\be
{1\over \a_{\rm eff}(M)} ={1\over \a } - 4p_1(M) = {1\over \a } - 4K'(M) \ .
\label{asfa}
\ee
This definition of $\a_{\rm eff}$  of course coincides with (\ref{runi}) when $N=2$ (it extends to any $N$).

In perturbation theory we obtain
\be
\langle e^{2\pi q a}\rangle =
1+3 \pi ^2 \alpha_{\rm eff}  q^2+\alpha ^2_{\rm eff} \frac{5 \pi ^4 q^4}{2}
+ \frac{1}{6} \pi ^2 \alpha ^3_{\rm eff} q^2 \left(7 \pi ^4 q^4-720 p_2\right)+O\left(\alpha ^4_{\rm eff}\right)\ .
\ee
which exhibits the dependence on the mass parameter $M$ through $ \alpha_{\rm eff}(M)$ and  $ p_2(M)$.


One can also consider an expansion where $\a$ is small but $q_*^2(M)\equiv q^2 \alpha_{\rm eff}(M) $ is an arbitrary fixed number.
We get
\be
\langle e^{2\pi q a}\rangle =
e^{\pi ^2 q_*^2} \left(1+2 \pi ^2 q_*^2-8\pi ^2 \alpha ^2_{\rm eff}(M)   p_2(M)  q_*^2 \left(2 \pi ^4 q_*^4+15 \pi ^2 q_*^2+15\right)+O\left(\alpha ^3_{\rm eff} \right)\right)\ .
\ee
The leading term is once again the  expectation value of the circular Wilson loop operator in ${\cal N}= 4$ SYM since in the strict $\a\to 0 $  limit with fixed $q_*$
the one-loop determinant is set to 1.

Let us apply these expressions to the study of the small and large $M$ behavior. 
At small $M$, one has $p_k(M)\approx c_k M^2 $ where 
$c_k=2k(2k+1)\zeta(2k+1)$.
We find 
\be
\langle e^{2\pi q a}\rangle =   e^{\pi ^2 q_*^2} \left(1+2 \pi ^2 q_*^2-160\pi ^2\zeta(5) \alpha ^2_{\rm eff}   M^2  q_*^2 \left(2 \pi ^4 q_*^4+15 \pi ^2 q_*^2+15\right) +O\left(M^4 \right)\right)\ ,
\label{quan}
\ee
with 
\be
q_*^2=q^2\a_{\rm eff}\ ,\ \ \ \ \a_{\rm eff}\approx \a + 24\zeta(3) \a^2 M^2\ .
\ee
We can read an $O(\a^2M^2)$ correction whose coefficient involves 
$\zeta(3) $ and $\zeta(5)$.

For large $M$ we use the following formulas
\bea
p_1(M) &=& \ell (M)+O(M^{-2}), \quad \ell (M)\equiv 2\log \big(M e^{\gamma+1} \big) \ ,
\nonumber\\
p_k(M) &=& 2\zeta(2k-1)+O(M^{-2(k-1)})\ ,\qquad k>1\ .
\eea
Thus we find 
\be
\langle e^{2\pi q a}\rangle =
e^{\pi ^2 q_*^2} \left(1+2 \pi ^2 q_*^2+O\left(M^{-2} \right)\right)\ .
\label{quas}
\ee
Now 
\be
{1\over \a_{\rm eff}} \approx {1\over \a } - 8\log (Me^{1+\gamma}) \ ,
\ee
which can be interpreted as the expectation value of the circular Wilson loop for ${\cal N}=2$ SYM written in terms of  the renormalized gauge coupling.

It should be noted that in the formulas  (\ref{quan}), (\ref{quas}) the additional $\a^3$ corrections are suppressed, either by $M^4$ --in the small $M$ limit--
or by $M^{-2}$ --in the large $M$ limit. Therefore these formulas capture the full $\a $ dependence if we restrict to the leading behavior in the power series expansion in the 
mass (or inverse mass) parameter.

\subsubsection{'t Hooft loop}

The expectation value of a supersymmetric  't Hooft loop operator carrying magnetic charge, with support on the equator of $S^4$,
is given in eq. (7.66) of \cite{Gomis:2011pf}. 
We set the $\theta $ parameter of the gauge coupling $\tau={\theta\over 2\pi}+{4\pi i\over g^2}$
 to $0$ and restrict to the zero instanton sector. The magnetic charges are $B=i{\rm diag}(p/2,-p/2)$ and, for concreteness, we shall consider the case of $p={\rm even}$. 
Using the relations
\be
\Gamma(z) G(z) = G(1+z)\ ,\ \ \ \ z\Gamma(z) = \Gamma(1+z)\ ,
\ee
we find the following  form for  expectation value of this ' t Hooft loop operator:
\bea
\langle T(p)\rangle &=& (Z^{{\cal N}=2^*})^{-1}\sum_{q=p,p-2,...,-p} \frac{p!}{ \left({p-q\over 2}\right)!\left({p+q\over 2}\right)!}\ e^{q^2\pi^2\over g^2}
\nonumber\\
&\times & \int da \  e^{-{16\pi^2 a^2\over g^2}  } \ \frac{ f_q(a,M)\ h_p(a,M) H^2(2a)}{H(2a-M)H(2a+M)}\ ,
\label{thooft}
\eea
where
\bea
h_p(a,M) &=& \left(\frac{\sinh(\pi (2a+M))\sinh(\pi (2a-M))}{\sinh^2(2\pi a)}\right)^{p\over 2}
\nonumber\\
&=& \left(\frac{\cosh(4\pi a) -\cosh(2\pi M)}   {2\sinh^2(2\pi a)}  \right)^{p\over 2}\ ,
\eea
$f_0(a)= 4 a^2$ 
and, for $q$ even, $|q|\geq 2$,
\be
f_q(a,M)=   \frac{ (4a^2)^{|q|\over 2}({q^2\over 4}+4a^2)}{ (4a^2- M^2)^{|q|\over 2}}
\prod_{k=1}^{{|q|\over 2}-1}\frac{ (k^2+4a^2)^{|q|-2k} }{
 ((k+iM)^2+4a^2)^{{|q|\over 2}-k} ((k-i M)^2+4a^2)^{{|q|\over 2}-k}}\ .
\ee

In particular,
\bea
f_2(a,M) &=& \frac{4 a^2 \left(1+4 a^2\right)}{4 a^2- M^2}\ ,
\nonumber\\
f_4(a,M) &=& \frac{ (4a^2)^2 \left(4 a^2+1\right)^2\left(4a^2+4\right) }{\left(4 a^2+(1-i M)^2\right) \left(4 a^2- M^2\right)^2 \left(4 a^2+(1+iM)^2\right)}\ .
\nonumber
\eea
These expressions show that for any finite magnetic charge $p$ the structure of perturbation series is the same, due to the fact that 
poles appear only at imaginary values of $a$, not on the real  axes.\footnote{Note that the poles at $2a=\pm M$ in $f_q(a,M)$ are canceled by
zeros of $h_p(a,M)$. 
This special point is related to the point identified in  \cite{Seiberg:1994aj},
 where a component of the adjoint hypermultiplet is massless.}
As in previous examples, the associated Borel series obtained by expanding the factor in the integrand multiplying $\exp[-a^2/\a]$ has finite radius of convergence, determined by the singularities
of the integrand in the complex plane. The integration over $a$ on each term of this series produces an additional factor $n!$.
The perturbative series for $\langle T(p)\rangle $, which is again an alternate series, therefore diverges like $n!$
but it is Borel summable because the integral over $a$ does not hit any singularity.

\medskip

For $M=0$, the expression for the 't Hooft loop operator  reduces to the one of ${\cal N}=4$ SYM theory, studied in \cite{Gomis:2009ir,Alday:2009fs,Drukker:2009id,Okuda:2010ke}.
Setting $M=0$, we see that $h_p\to 1$ and $f_q$ simplifies:
\be
 f_q(a,M=0)={q^2\over 4}+4a^2\ .
\ee
We get ($p=2m$)
\bea
\langle T(p)\rangle_{M=0} &=& (Z^{{\cal N}=4})^{-1}\sum_{n=-m}^m  
\left( \begin{matrix}
2m \\
m+n
\end{matrix}\right)
e^{4\pi^2n^2\over g^2} \int da \  e^{-{16\pi^2a^2\over g^2} } \big(n^2+4a^2\big)
\nonumber\\
&= & \sum_{n=-m}^m \left( \begin{matrix}
2m \\
m+n
\end{matrix}\right)
\ e^{4\pi^2n^2\over g^2}\ \left(1+\frac{8\pi^2 n^2}{g^2}\right)\ .
\label{th4}
\eea
Under  an $S$ duality transformation $\tau\to -1/\tau$, 't Hooft loops are mapped to Wilson loops.
The precise relationship is discussed in \cite{Gomis:2011pf}.
{}For $\theta=0$, the $S$ duality  transformation is $g\to 16\pi^2/g$ and a given term in (\ref{th4})   is mapped to the ${\cal N}=4$ result (\ref{adru}) with $\ell=n $.


\medskip

 The expectation value of the 't Hooft loop operator
determines the behavior of the monopole-antimonopole potential.
The large loop length behavior of the 't Hooft loop operator is believed to indicate whether the theory is in a confining, Higgs or Coulomb phase. 
For a confining phase, the 't Hooft loop operator is expected to exhibit a perimeter law, whereas for the Higgs phase it should exhibit an area law. 
In the present context the 't Hooft loop is placed around the equator of $S^4$ and it has therefore a length
proportional to $r$, the $S^4$ radius.
{} Although the scale of the 't~Hooft loop cannot be disentangled from 
the scale of  $S^4$, it is nevertheless interesting to examine its behavior
as $r$ is changed  with respect to the mass parameter.
In this manner by  large 't Hooft loop one means $r\gg 1/M$ or, since we have set $r=1$, it means looking
at the behavior at $M\gg 1$.

 Let us first expand $\langle T(p)\rangle$ in powers of $g $ (here we only study the $g\ll 1 $ regime, since we are ignoring instanton contributions). 
At $g\ll 1$, in the sum over $q$, the terms with $q=\pm p$ dominate because of the factor
$e^{\pi^2 q^2/g^2}$. In order to study the behavior at $M\gg 1$ it is convenient to express the one-loop determinant factor in terms of $B^{{\cal N}=2^*}(a^2,M)$
defined in (\ref{hatB}).
In particular, for $p=2$, we need to compute
\be
\langle T(p=2)\rangle =\frac{2e^{4\pi^2\over g^2}   }{Z^{{\cal N}=2^*}}
\int da \ (1+4a^2) \frac{4a^2 (\cosh(4\pi a) -\cosh(2\pi M))}{2\sinh^2 (2\pi a)(4a^2-M^2)} 
 \ e^{-{a^2\over \a_{\rm eff}}} \hat B^{{\cal N}=2^*}(a^2,M)
\label{thoof}
\ee
$\a_{\rm eff}$ has been defined in (\ref{asfa}).
We find
\be
\langle T(p=2)\rangle =2e^{4\pi^2\over g^2} \big(T^{(1)}+\cosh(2\pi M)\ T^{(2)}\big)\ ,
\ee
with
\bea
T^{(1)} &=& -\frac{1}{4\pi ^2 M^2 \alpha_{\rm eff}  }-\frac{3+\big(3+5 \pi ^2\big) M^2}{6 \pi ^2 M^4}
\nonumber\\
&-&\frac{\alpha_{\rm eff} }{5\pi ^2
   M^6}  \left(30 M^4 p_2+\pi ^2 \left(25+\pi ^2\right) M^4+5 \left(3+5 \pi ^2\right) M^2+15\right)+O\big(\alpha ^2\big)
\nonumber\\
T^{(2)}  &=& \frac{1}{4 \pi ^2  M^2  \alpha_{\rm eff}}+\frac{3-\big(\pi ^2-3\big) M^2}{6 \pi ^2 M^4}
\nonumber\\
&+& \frac{\alpha_{\rm eff}}{5 \pi ^2 M^6}  \left(30 M^4 p_2+\pi ^2 \left(\pi ^2-5\right) M^4-5 \left(\pi ^2-3\right) M^2+15\right)+O\big(\alpha ^2\big)\ .
\eea
The origin of the $1/\a_{\rm eff}$ behavior in the leading term can be traced back to  the fact that the integrand in the numerator now contains a term $4a^2+q^2/4=4a^2+1$, whereas in the denominator $Z^{{\cal N}=2^*}$ contains the usual 
Vandermonde factor $4a^2$.

In the large $M$ limit we therefore find the following behavior:
\be
\langle T(p=2)\rangle =4 e^{4\pi^2\over g^2} \frac{e^{2\pi M}}{g^2 M^2}
\big(1-\frac{ g^2}{2\pi^2} \log M+O(g^4 )+O(M^{-2})\big)\ .
\ee
This is readily extended to  general $p$.  For the dominant $q=\pm p$ term at $g\ll 1$ we  now find
\be
\langle T(p)\rangle =c(p)\ \frac{1}{ \left(e^{-2\pi {\rm Im}(\tau)} M^4\right)^{p^2\over 8}}  \ \frac{ e^{\pi p r M}}{g^2} \ \left(1-\frac{ g^2}{2\pi^2} \log(M)+O(g^4 )+O(M^{-2})\right), \qquad Mr\gg 1\ ,
\label{TTT}
\ee
where $\tau= {\theta\over 2\pi} + {4\pi i\over g^2}\to {4\pi i\over g^2}$ and we have restored the $r$ dependence. 

In \cite{Seiberg:1994aj}, for the theory on ${\bf R}^4$,  it was pointed out that the limit to pure $N=2$ SYM theory is achieved from the $N=2^*$ SYM theory  by taking $M$ large and holding $e^{2\pi i \tau }M^4$ fixed. 
Strikingly, in (\ref{TTT}) $\exp(2\pi i\tau )$ appears in the right combination with the  power of $M$.
We now see that in our context the prescription by  Seiberg and Witten has a simple physical explanation.
The coupling constant dependence must be through the renormalized gauge coupling, 
\be
{1\over g_R^2}={1\over g^2 }\left(1-\frac{ g^2}{2\pi^2} \log(M)\right)\ .
\ee
At large $M$, there should not be any other dependence on the bare gauge coupling which is not through the combination forming $g_R$. Indeed,
(\ref{TTT}) can be written as
\be
\langle T(p)\rangle =c(p)\ \frac{1}{ \left(e^{-2\pi {\rm Im}(\tau_{\rm R})} \right)^{p^2\over 8}}  \ \frac{ e^{\pi p r M}}{g^2_{\rm R}} \ \left(1+O(g^4 )+O(M^{-2})\right), \qquad Mr\gg 1\ ,
\label{TTTk}
\ee
The remaining dependence on $M$ is in the exponential factor $e^{\pi p r M}$. This originates from $h_p$, which in turn comes from
two contributions, the one-loop determinant at the equator and a contribution from
monopole field configurations that screen the monopole charge $B$. Viewing $N=2^*$ SYM theory as pure $N=2$ SYM with a UV cutoff $M$, this factor
might be indicating a very singular behavior of the expectation value of the 't Hooft loop operator in the UV limit, or it might  be just reflecting the fact that the loop becomes very
long and it follows a perimeter law. To clarify the meaning of this behavior it would be useful to compute the expectation value of a 't Hooft loop operator with support on a circle of arbitrary length $L$. This
would allow one to consider a limit $Mr\gg 1$ at fixed $L$.




\section{Concluding remarks}

One of the lessons of the  matrix model examples studied here is that  the divergence of the perturbation series is not inherent
to the theory but rather
 produced by an  illegitimate step in the Feynman-diagram evaluation of functional integrals,
when  commuting an integral with a sum: the power expansion of vertex operators with the functional integration. 
In this respect, the Borel summation  emerges not as an artificial
trick ``to make convergent what is actually divergent", but it is rather built in the theory. 
In the simplest $SU(2)$ cases the exact gauge theory partition function  automatically appears
in the form of a Borel transform.
It is tempting to conjecture that  gauge theories are free from divergent sum pathologies  if care is taken when commuting 
the functional integral with other integrals or infinite sums.\footnote{For example, in the $U(2)\times U(2)$ ABJM model, one can compute the integral (\ref{jorg})
by expanding ${\cal S}(v/k)$ in power series in $v/k$ --which would lead to a divergent series -- or  by residues \cite{Okuyama:2011su}
-- which would lead to a finite sum of terms.}
More precisely,  for certain observables the Borel function may have poles on the real axes. In this case the Borel technique requires deforming the 
integration contour,
a procedure that is not free from ambiguities and typically produces imaginary parts.
Thus, a more precise conjecture is that there should be  no divergent sum pathologies in observables where an imaginary part is not expected.
For observables where an imaginary part is expected (like e.g. two-point functions of  unstable particles) a suitable
contour deformation prescription may not only render the integral convergent but also give rise to the expected imaginary part.


\medskip

To conclude, we summarize the main results.

\begin{itemize}

\item The perturbation series for the partition function in the $U(N)\times U(N)$ ABJM theory is an alternate asymptotic series with
$(-1)^m (2m)! (2/\pi k)^{2m}$ asymptotic behavior. This behavior is not associated with instantons (which are absent in three dimensions) but with poles on the complex plane
of the one-loop determinant.
The integral defining the partition function is convergent, so the divergent perturbation series is Borel summable.
For the $U(2)\times U(2)$ theory, a simple integral representation illustrates the origin of the  divergent perturbation series.

\item For ${\cal N}=2$  $SU(2)$ SYM theory with four massless hypermultiplets, we computed the asymptotic behavior of the perturbation series
for the partition function and Wilson loop operator, in the zero and one instanton sector, and again found a Borel summable perturbation series, with asymptotic behavior $(-1)^n n! \a^{n} $,\ $\a=g^2/16\pi^2$.

\item For the ${\cal N}=2^*$ theory in the planar limit, the Wilson loop at weak coupling has the same form as in  ${\cal N}=4$  SYM theory upon replacing the 't Hooft coupling $\lambda $
by 
$$
\lambda_{\rm eff}(M)= \lambda \ \left(1+ \frac{\lambda K'(M)}{8\pi^2}\right)\ ,\qquad \frac{\lambda K'(M)}{8\pi^2}\ll 1\ .
$$
At strong coupling and small mass parameter, we find $W(C_{\rm circle})\approx e^{2\pi T} $ with
$$
T= \frac{\sqrt{\lambda}}{2\pi} \left(1+\frac{M^2}{2}\right)\ ,\qquad \lambda\gg1,\ \ M\ll 1\ ,
$$
where $T$ can be viewed as the tension of a semiclassical string configuration.
In all cases   augmenting the mass parameter produces an increase of  $W(C_{\rm circle})$.
Finally, we have examined the expectation value $\langle T(p)\rangle $ of a  circular 't Hooft loop operator of magnetic charge $p/2$ in the $SU(2)$ theory.
Like for the Wilson loop, the associated perturbation series in the zero instanton sector is divergent with a $(-1)^n n!$ asymptotic behavior,
and Borel summable. At large masses, where the theory is expected to flow to pure  ${\cal N}=2$  $SU(2)$ gauge theory with a UV cutoff $M$,
 $\langle T(p)\rangle $ seems to increase as $e^{\pi p r M}$ in the weak coupling regime.

\end{itemize}

It would be interesting to study sectors of higher instanton charge in ${\cal N}=2$ $SU(2)$ SYM  theories 
to see  if more complicated instanton configurations may give rise to singularities in the Borel function lying on the integration region. This could provide
an arena to clarify issues of ambiguities in contour deformation in the context of non-abelian gauge theories from closed analytic expressions.

It would also be interesting to compute observables in $N=2^*$ $SU(2)$ SYM theory with a characteristic length $L$.
By introducing  a scale one can separate UV and IR physics in the series expansions and, in particular, 
see if the use of perturbation theory is problematic in the IR region, like in the case of QCD renormalons.


\section*{Acknowledgements}

The author would like to thank Jaume Gomis and Mithat \" Unsal for useful discussions
and the  Perimeter Institute for Theoretical Physics for hospitality.
He acknowledges support by MCYT Research
Grant No.   FPA 2010-20807 and Generalitat de Catalunya under project 2009SGR502.

\end{document}